\documentstyle[times,pramana,epsf,floats]{ias}
\begin{document}
\mark{{Two problems in thermal field theory}{Fran\c cois Gelis}} \title{Two problems
  in thermal field theory\footnote{Talk given at the 6th Workshop on
    High Energy Particle Physics, Jan. 3-15 2000, Chennai, India}}

\author{Fran\c cois Gelis} 
\address{Brookhaven National Laboratory,
  Nuclear Theory, Bldg 510A, Upton, NY-11973, USA} 

\abstract{ In this talk, I review recent progress made in two areas
  of thermal field theory. In particular, I discuss various approaches
  for the calculation of the quark gluon plasma thermodynamical
  properties, and the problem of its photon production rate.}

\maketitle
\section{Introduction}
At low temperature and density, quarks and gluons appear only as
constituents of hadrons because of confinement. However, QCD lattice
simulations predict that above a certain temperature or density, nuclear
matter could undergo a phase transition after which quarks and gluons
are deconfined from hadrons, and form a new state of matter called
``quark-gluon plasma''.
\begin{figure}[htbp]
\centerline{\hfill
\epsfysize=4cm
\epsfbox{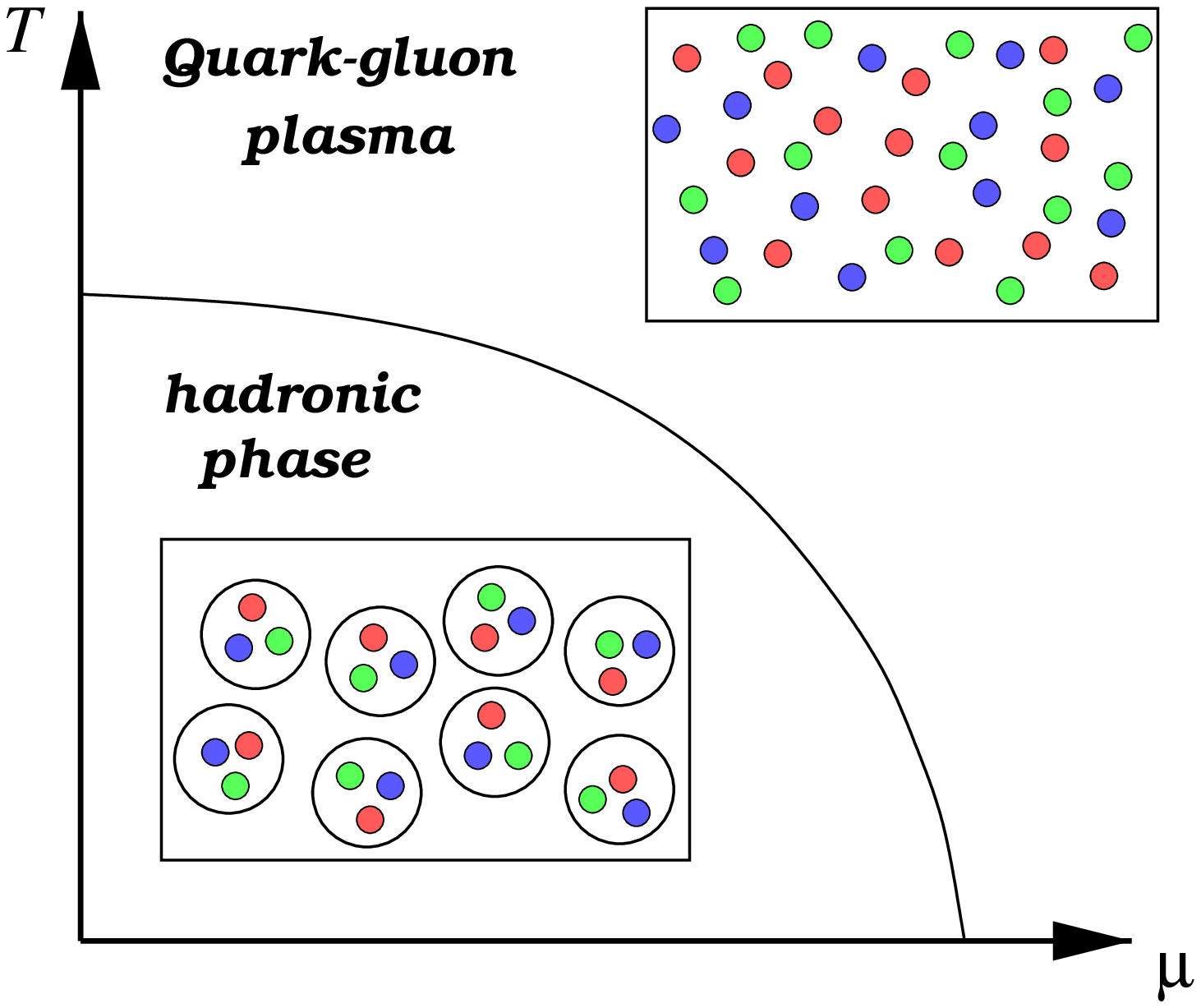}\hfill\hfill
\epsfysize=3cm
\epsfbox{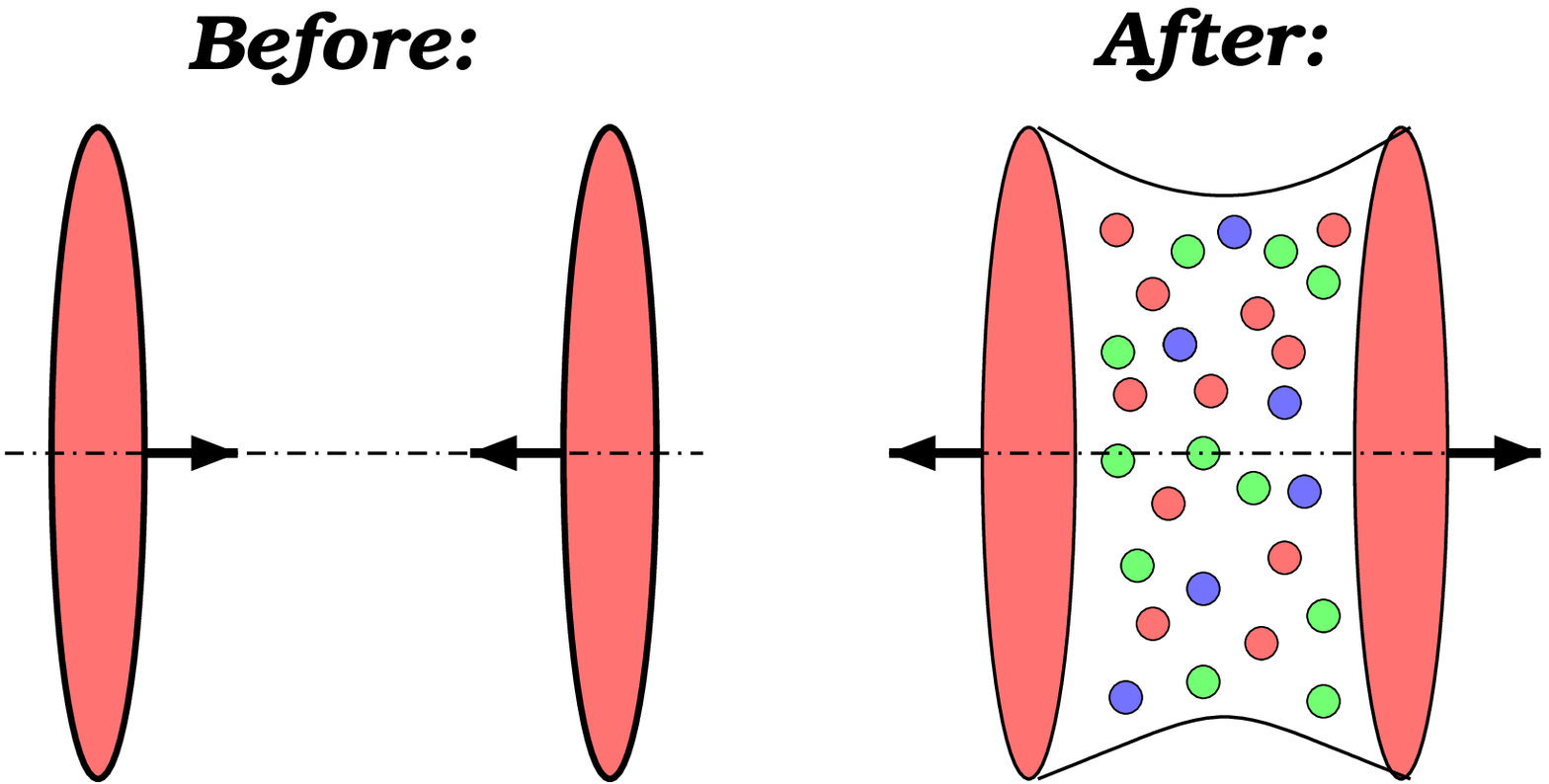}\hfill}
\caption{Left: simplified phase diagram for QCD. 
  Right: a heavy ion collision.}
\label{fig:transition}
\end{figure}
In order to create in the laboratory the conditions of temperature and
density necessary for this transition, heavy nuclei are collided at
very high energies. To be able to detect the formation of the
quark-gluon plasma in such collisions, one needs to find observables
that are sufficiently different in a QGP and in hot hadronic matter.

Thermal field theory is one of the possible tools for studying the
plasma phase. Basically, thermal field theory is the formalism
obtained by merging quantum field theory and the tools of statistical
mechanics: its only difference with zero temperature field theory is
that it incorporates a statistical ensemble of particles in the
system, and that those particles can participate in reactions. More
formally, the statistical ensemble appears in the definition of
thermal Green's functions: at statistical equilibrium, the thermal
average of an operator $A$ is defined as ${\rm Tr}(e^{-H/T}A)/Z$
instead of $\left<0|A|0\right>$. In order to calculate perturbatively
these thermal Green's functions, there is a set of Feynman rules very
similar to the zero temperature ones\cite{Frules}.

An improvement over the bare thermal perturbative expansion has been
proposed in the early nineties by \cite{HTL}, which amounts to the
resummation of 1-loop thermal corrections. Indeed, it was realized
that these loop corrections (known as Hard Thermal Loops) are of the
same order of magnitude as the corresponding tree-level amplitude when
the external momenta are soft.  In this context, {\sl hard} refers to
momenta of order the temperature $T$, while {\sl soft} means of order
$gT$, where $g$ is the coupling constant of the theory. The
resummation of HTLs can be seen as a reordering of the perturbative
expansion.

\begin{figure}[htbp]
\centerline{\hfill
\epsfysize=2.7cm
\epsfbox{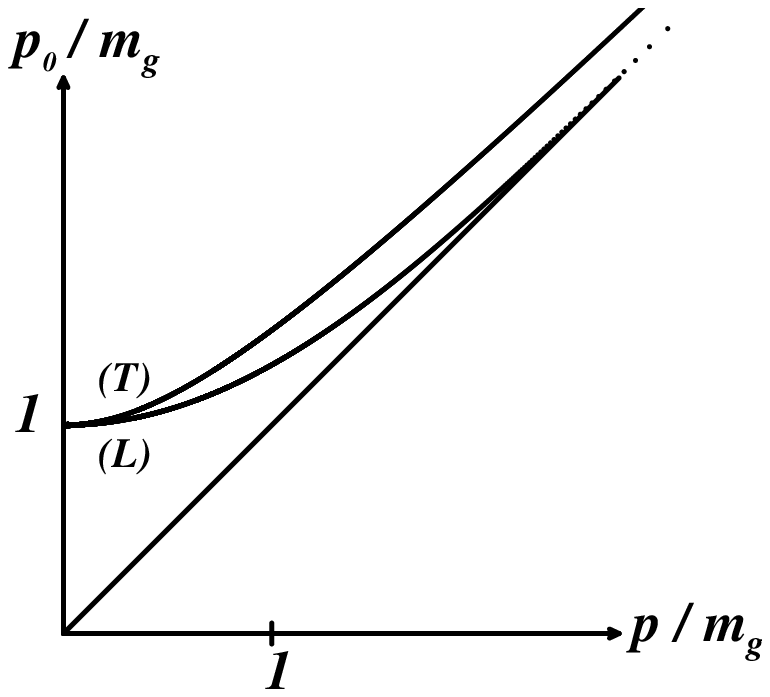}\hfill
\epsfysize=2.7cm
\epsfbox{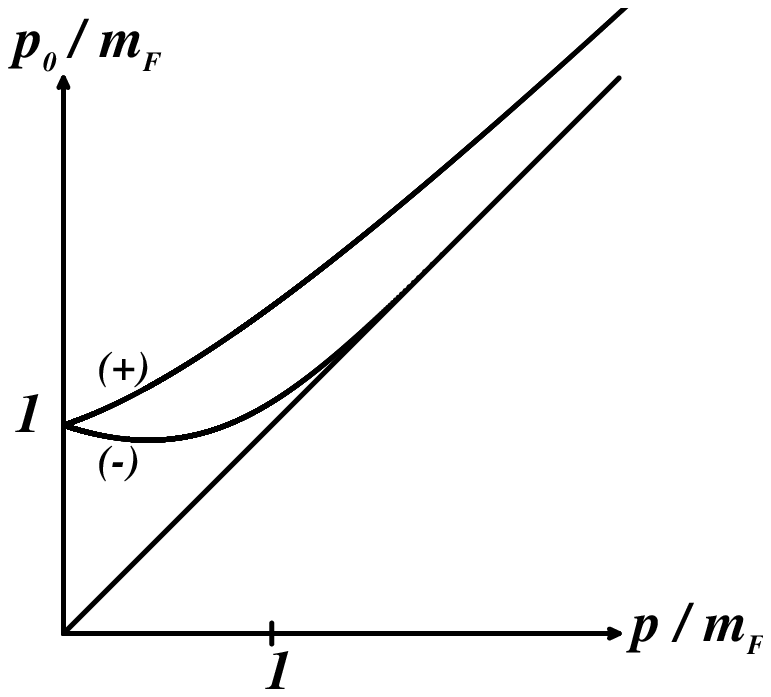}\hfill\hfill
\epsfxsize=5cm
\epsfbox{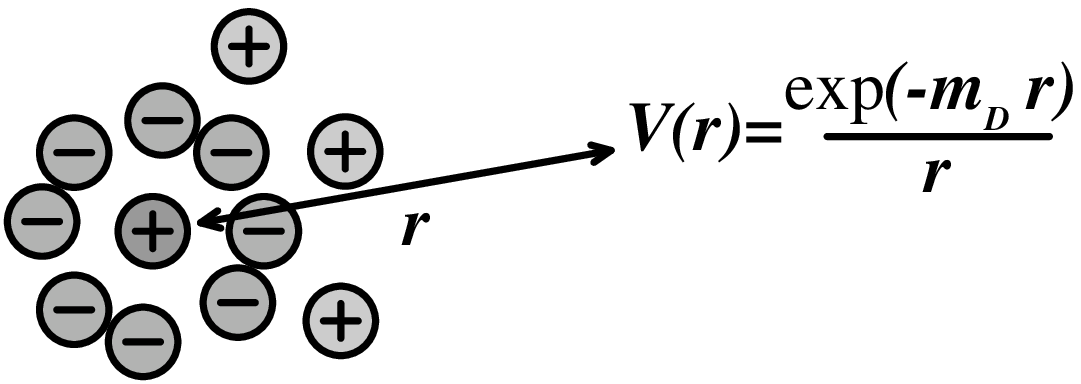}\hfill}
\caption{Left: gluon dispersion curves. Middle: quark 
  dispersion curves ($m_g$ and $m_{_{F}}$ are thermal masses of order
  $gT$). Right: Debye screening of the electric field.}
\label{fig:HTL}
\end{figure}

Physically, this resummation modifies the dispersion
curve of elementary excitations of the theory (see Fig.~\ref{fig:HTL}
for QCD) by providing them a thermal mass of order $gT$. Additionally,
in the static limit ($p_0\to 0$) in the space-like region, the
longitudinal gauge boson has a non vanishing self-energy $m_{_{D}}^2\sim
g^2 T^2$ which provides a screening of static electric fields in the
plasma.
This phenomenon is known as Debye screening (see the illustration on
Fig.~\ref{fig:HTL}) and is due to the fact that a test charge placed
in the plasma polarizes the medium around it, so that the Coulomb
field it creates at large distance is exponentially suppressed. At
this level of approximation, there is no screening for static magnetic
fields\footnote{In QED, this result is known to hold at all orders of
  perturbation theory. However, this proof cannot be generalized to
  QCD. In fact, a magnetic screening mass is expected for QCD at the
  non-perturbative scale $g^2T$.}. Finally, the resummation of hard
thermal loops also includes in the effective theory the Landau
damping, coming from the fact that the quark and gluon self energies
have an imaginary part in the space-like region. Note also that these
self-energies are purely real in the time-like region, which means
that the elementary excitations are stable in this framework. Their
decay is a next-to-leading effect, and their lifetime is of order
$(g^2T)^{-1}$.

\section{Thermodynamics of a quark-gluon plasma}
\subsection{Convergence of the perturbative expansion}
A lot of work has been devoted in the past years to the calculation of
thermodynamical properties of a quark gluon plasma, and in particular
of its pressure. The pressure can be obtained by differentiation with
respect to the volume from the free energy, which is itself given by
$F\equiv \ln{\rm Tr}\,(\exp(-H/T))$. $F$ can be calculated
perturbatively as a sum of vacuum diagrams. Although this problem
looks straightforward, early attempts showed that the bare
perturbative expansion of this quantity has a very poor convergence
\cite{BraatN,ArnolZ1}.  \catcode`\@=11 \global\advance\c@figure by 1
\catcode`\@=12
\begin{figure}[htbp]
  \centerline{\hfill\parbox{6cm}{\epsfysize=4.2cm
      \epsfbox{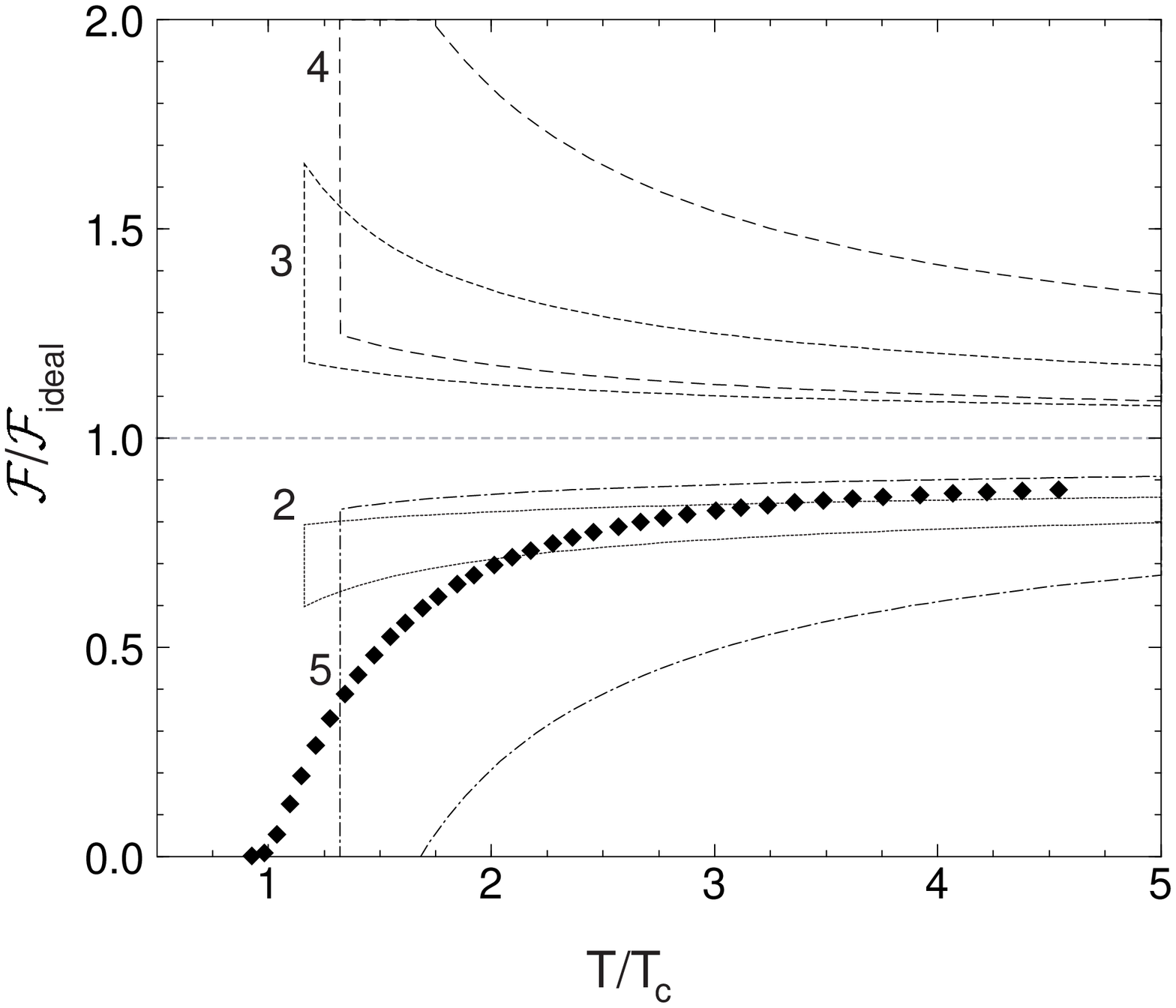}}\hfill \parbox{5.5cm}{\small{{\bf
          Figure\ \ 3.\ \ }Normalized QCD (without quarks) free energy
        as a function of temperature \cite{BraatN}. The dotted line is
        a result from lattice QCD \cite{BoydA1}. The various curves
        come from perturbation theory respectively to order $g^2$,
        $g^3$, $g^4$, and $g^5$. The bands correspond to a variation
        of the renormalization scale by a factor $2$. $T_c$ is the
        critical temperature.}} \hfill }
\label{fig:qcd-free-energy}
\end{figure}
Practically, the coefficients in front of higher orders in $g$
were found to be quite large, limiting the applicability of these
results to very small values of the coupling constant. This is
illustrated for QCD on Fig.~3.

\subsection{Screened perturbation theory in scalar theories}
In order to overcome this difficulty, various resummation schemes have
been proposed to improve the convergence of the perturbative
expansion. Let me first discuss some of those strategies in the case
of a scalar field theory with a quartic coupling $g^2\phi^4$, since
this model illustrates the methods while avoiding complications
related to gauge invariance. In this model also, the convergence of
the bare perturbative expansion is very poor \cite{ParwaS2}.

A first method used to solve this problem is known as ``screened
perturbation theory'', and amounts to reorganize the perturbative
expansion by inserting a fictitious mass term $m$ in the Lagrangian
\cite{KarscPP1}:
\begin{equation}
{\cal L}={1\over 2}\partial_\mu\phi\partial^\mu\phi -{1\over 2}m^2\phi^2 +{{g^2}\over{4!}}\phi^4+{1\over 2}m^2\phi^2\; .
\end{equation}
The idea behind this improvement is that medium effects tend to give a
mass to elementary excitations: in other words, some of the
interactions can be hidden in an effective mass. In this framework,
one has a massive propagator $(P^2-m^2)^{-1}$, and an additional
interaction vertex proportional to $m^2\phi^2$. Of course, since this
method adds and subtracts a mass term, exact results do not depend on
the parameter $m$. In practice, since one truncates the perturbative
expansion at a given finite order, there is a residual dependence upon
$m$ in the result.  Therefore, one has to choose the value of this
mass. Some possibilities are listed below:

\noindent (1) One can use the perturbative value for the thermal 
mass in this theory. At 1-loop, that amounts to choose $m^2=g^2
T^2/24$, and this choice is equivalent to the resummation of HTLs.

\noindent (2) One can get this mass from a gap equation.

\noindent (3) Another possibility is to argue that since this mass 
term is not a physical quantity, one should choose $m$ in order to
minimize the sensitivity of the result on $m$.

These three methods give the same $m$ at small $g$, but may differ for
larger values of the coupling constant. The authors of \cite{KarscPP1}
found that this reorganization of the perturbative expansion lead to
significant improvements of its convergence, as illustrated on
Fig.~4.  \catcode`\@=11 \global\advance\c@figure by 1 \catcode`\@=12
\begin{figure}[htbp]
  \centerline{\hfill\parbox{6cm}{\epsfysize=4.2cm
      \epsfbox{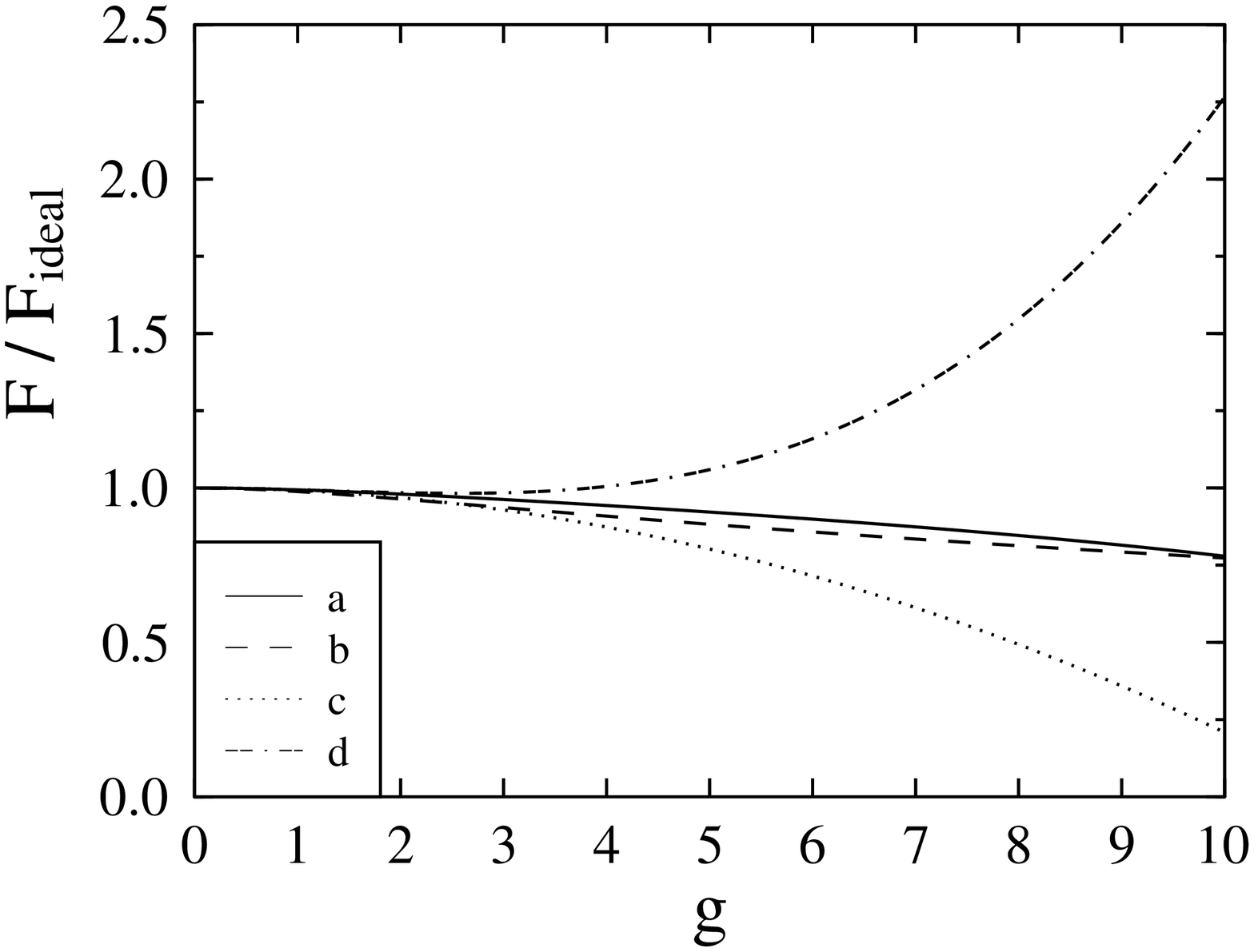}}\hfill \parbox{5.5cm}{\small{{\bf
          Figure\ 4.\ \ }Results of the screened perturbative
        expansion for the free energy as a function of the coupling
        constant in scalar field theory \cite{KarscPP1}. (a) and (b):
        first two orders in screened perturbative expansion. (c) and
        (d): first two orders in the bare perturbative expansion
        \cite{ParwaS2}.}} \hfill }
\label{fig:screened-pert}
\end{figure}

The idea that it may be enough to tune the mass of quasi-particles to
obtain a good estimate of the thermodynamical properties of the system
has also been explored for QCD in \cite{PeshiKP1}. The authors of this
work show that it is possible to reproduce the lattice results for the
pressure of a $SU(3)$ Yang-Mills gas just by introducing a mass in the
propagator of gluons. Physically, the fact that the pressure goes down
when $T\to T_c^+$ is a sign that partons feel a strong attractive
force, and can be seen as a precursor sign of confinement. In order to
reproduce this behavior, the mass introduced in \cite{PeshiKP1} has to
increase when $T\to T_c^+$, mimicking the fact that partons are
trapped in heavy bound states.

\subsection{Consistent approximation schemes}
In addition to the fact that the applicability of the above ansatz to
gauge theories is not obvious (it cannot be applied at higher orders
in QCD because it lacks vertex corrections for gauge invariance), it
also suffers from the fact that there are many ways to choose the mass
parameter. A generalization of this method that overcomes the second
of those problems can be derived by making use of a formula derived by
Luttinger and Ward \cite{LuttiW1} for the thermodynamical potential.
In the case of a scalar theory, this formula reads \cite{PeshiKP2}:
\begin{equation}
\Omega={1\over 2}TV\int\!\!\!\!\!\!\!\!\!\sum \left\{
\ln(-\Delta^{-1})+\Delta\Pi
 \right\}+\Omega^\prime\; ,\,
\Omega^\prime=-\sum\limits_{n}{1\over{4n}}TV
\int\!\!\!\!\!\!\!\!\!\sum \Delta\Pi_{n}\; ,
\end{equation}
where $\Delta$ is the propagator, $\Pi\equiv\Delta_o^{-1}-\Delta^{-1}$
the self-energy, and where the sum in the second line is a skeleton
expansion containing only two particle irreducible vacuum
contributions. This formula must be supplemented by the following
variational principle:
\begin{equation}
{{\delta\Omega}/{\delta\Delta}}=0\; ,
\label{eq:consistency}
\end{equation}
which states that the thermodynamical potential should be stationary.
This condition ensures thermodynamic consistency, and gives a relation
between the self-energy and the propagator by $TV
\Pi=-2\delta\Omega^\prime/\delta\Delta$. Within this framework, one
can derive approximation schemes that preserve consistency. Indeed,
one can truncate the skeleton expansion of $\Omega^\prime$, apply
Eq.~({\ref{eq:consistency}}) to relate the propagator and the
self-energy, and then compute the resulting $\Omega$. For instance, if
one takes the following approximation for $\Omega^\prime$:
\setbox1=\hbox to 1.7cm{\epsfxsize=1.7cm\epsfbox{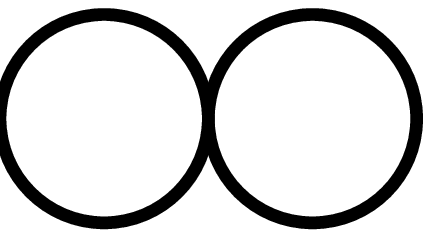}}
\setbox2=\hbox to 1.7cm{\epsfxsize=1.7cm\epsfbox{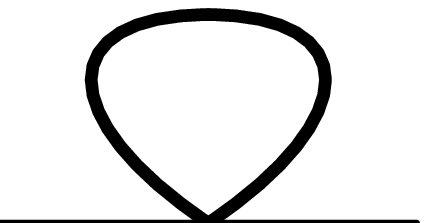}}
\begin{equation}
\Omega^\prime_1=-{{TV}\over 4}\;\raise -4mm\box1\; ,\quad
{\rm then\ one\ gets:\ \ \ }
\Pi=\;\raise -2mm\box2\; .
\end{equation}
Therefore, at this level of approximation, this ansatz is equivalent
to the screened perturbation theory where the mass $m^2$ is determined
by a gap equation. As a consequence, results obtained at this level of
approximation \cite{PeshiKP2} are very similar to those displayed in
Fig.~4.

At higher orders in the skeleton expansion, one would obtain
non-constant self-energies with a much richer analytic structure:
keeping higher orders for $\Omega^\prime$ has the effect to enlarge
the functional space in which the minimum of $\Omega$ is searched and
therefore increases the chances to find a better minimum, closer to
the absolute minimum that gives the exact $\Omega$.

This approximation scheme can in principle be generalized to gauge
theories, where $\Omega$ should be minimized with respect to the
2-point function, but also with respect to variations of vertices
\cite{FreedM} (in order to preserve gauge invariance). Although
possible in principle, this extended variational principle is very
difficult to use due to its complexity. Instead of that, attempts have
been made in which one trades the consistency for gauge invariance
\cite{QCD-consist}: the gauge invariance is recovered by using the HTL
approximation for propagators and self-energies, but the consistency
is spoiled because HTLs are not exact solutions of the consistency
equation.

\subsection{HTL perturbative expansion}
The spirit of this method\cite{AnderBS} is very similar to the
screened perturbation theory, applied to the case of QCD. One adds and
subtract to the QCD lagrangian the term $m^2{\cal L}_{_{HTL}}$ that
generates hard thermal loops:
\begin{equation}
{\cal L}={\cal L}+m^2{\cal L}_{_{HTL}}-m^2{\cal L}_{_{HTL}}\; ,
\end{equation}
where I have made explicit the fact that the HTL lagrangian is
proportional to a thermal mass $m^2$. The subtracted part is treated
as counterterms. Again, since the full Lagrangian does not depend on
$m^2$, one is free to choose $m^2$ at will. One possibility is to use
the perturbative value $m^2\sim g^2 T^2$ for this parameter
\cite{HTL}. At 1-loop order, this gives the result displayed in
Fig.~5.  \catcode`\@=11 \global\advance\c@figure by 1 \catcode`\@=12
\begin{figure}[htbp]
  \centerline{\hfill\parbox{6cm}{\epsfysize=4.2cm
      \epsfbox{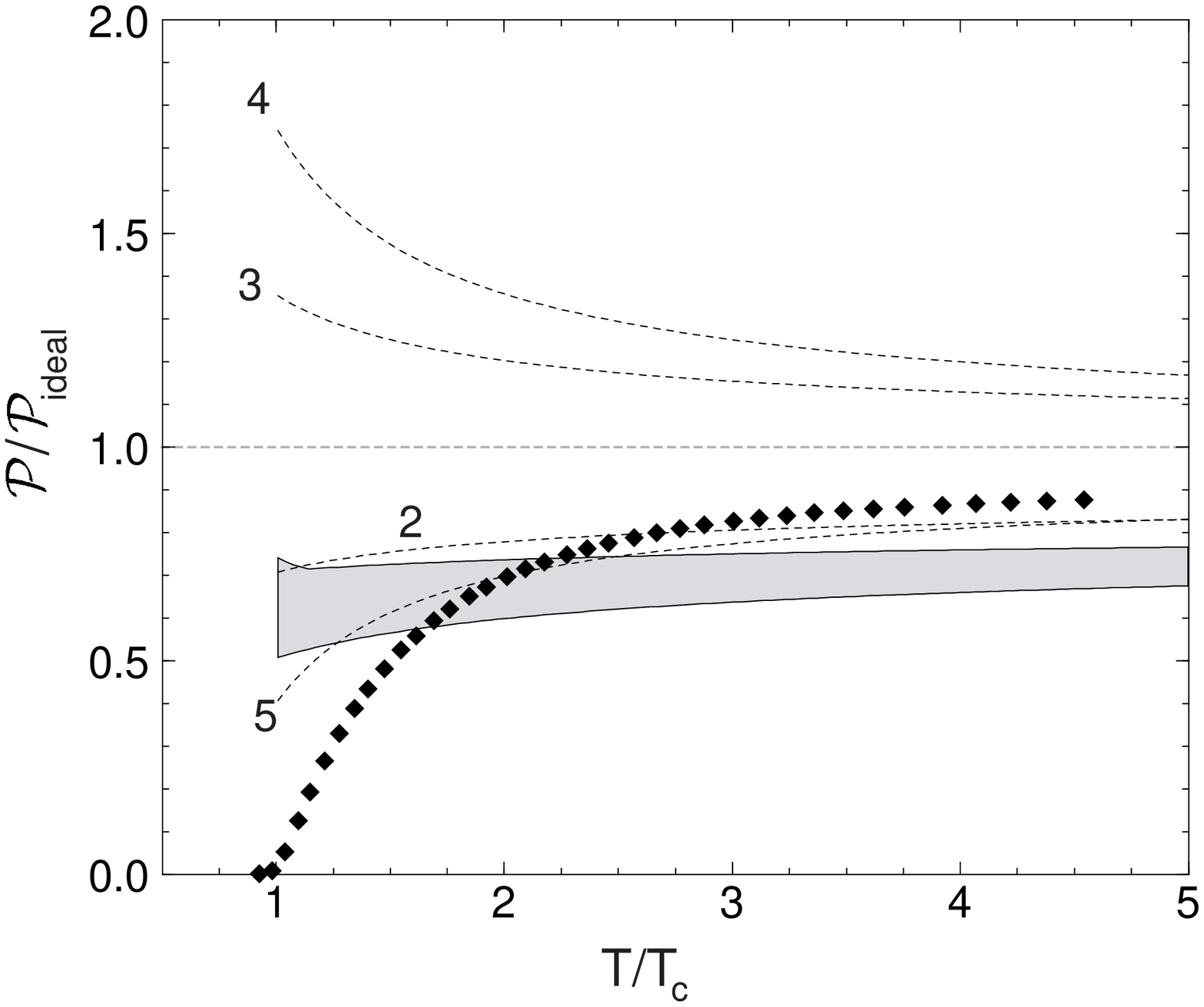}}\hfill
    \parbox{5.5cm}{\small{{\bf Figure\ 5.\ \ }Results of the 1-loop
        HTL perturbative expansion (shaded area) for the normalized
        pressure \cite{AnderBS}. Dotted curve: lattice data from
        \cite{BoydA1}. 2,3,4,5: 2nd, 3rd, 4th and 5th order results in
        the bare perturbative expansion\cite{BraatN}.}} \hfill }
\label{fig:HTL-resummation}
\end{figure}

At higher orders, it is possible to determine $m^2$ in order to
minimize the free energy with respect to $m^2$. This is very similar
to the variational principle discussed above, restricted to the
1-dimensional sub-manifold spanned by the HTL propagator. This
restriction tremendously simplifies the variational principle, while
maintaining exact gauge invariance (because ${{\cal L}_{_{HTL}}}$ is
gauge invariant). Obviously, this method will give good results if
(and only if) this 1-dimensional sub-manifold gets close to the absolute
minimum. In other words, it will work if the HTLs include the relevant
physics of the QGP thermodynamics...


\section{Photon and dilepton production}
\subsection{1-loop calculation}
Another physical quantity of interest for a quark gluon plasma is its
photon production rate. Indeed, the size of the QGP expected in nuclei
collisions is smaller than the photon mean free path (photons are
weakly coupled). As a consequence, a photon produced in the plasma can
escape and be detected without undergoing other interactions.  Photons
and dileptons are therefore very clean probes of the state of the
matter at the time they were emitted. In the following, I focus mainly
on photons with a small invariant mass, since this is the region where
the calculation is the most sensitive to problems like infrared and
collinear singularities.

\catcode`\@=11 \global\advance\c@figure by 1
\catcode`\@=12
\begin{figure}[htbp]
  \centerline{\hfill\parbox{5.5cm}{\epsfxsize=5.5cm
      \epsfbox{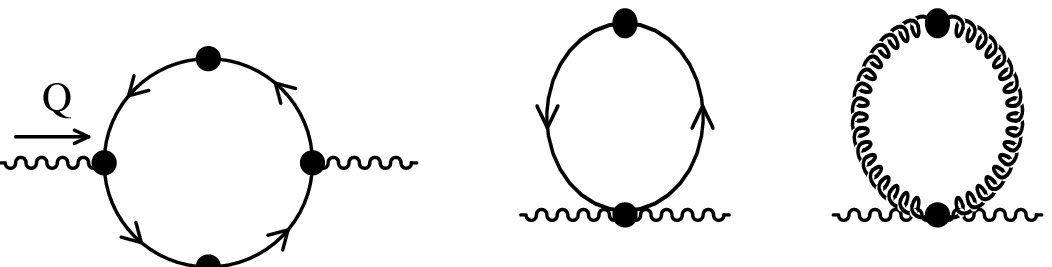}}\hfill \parbox{5.5cm}{\small{{\bf
          Figure\ 6.\ \ } 1-loop diagrams for the photon polarization
        tensor in the HTL effective perturbative expansion.}} \hfill }
\label{fig:1-loop}
\end{figure}
In thermal field theory, the photon/dilepton rate is proportional to
the imaginary part of the photon polarization tensor ${\rm
  Im}\,\Pi{}^\mu{}_\mu$ \cite{rate}. In the HTL effective theory, the
1-loop contributions to this 2-point function have been evaluated both
for soft \cite{soft-gamma} and hard \cite{hard-gamma} real photons.
The 1-loop diagrams are given in Fig.~6. For hard photons, one can
neglect the HTL effective vertices, and the result was found to be of
order ${\rm Im}\,\Pi{}^\mu{}_\mu(Q)\sim e^2 g^2 T^2$. For soft
photons, it was found to be of order ${\rm Im}\,\Pi{}^\mu{}_\mu(Q)\sim
e^2 g^4 T^3/q_0$. In addition, the latter contribution has a collinear
singularity that, after being regularized \cite{FlechR1} by a thermal
mass of order $gT$, gives an extra factor $\ln(1/g)$.  For the
contribution to soft photon production, one can note that the 1-loop
result is extremely suppressed (it behaves like $g^4$), because these
1-loop diagrams have a very small phase space since the loop is soft.
Another problem with this 1-loop calculation (both for hard and soft
photons) is that bremsstrahlung does not seem to contribute, contrary
to results known from classical plasma physics.

\subsection{2-loop calculation}
In fact, one can check that the above two problems are solved if one
takes into account the 2-loop contributions of Fig.~7. 
\catcode`\@=11 \global\advance\c@figure by 1
\catcode`\@=12
\begin{figure}[htbp]
  \centerline{\hfill\parbox{5.5cm}{\epsfxsize=5.5cm
      \epsfbox{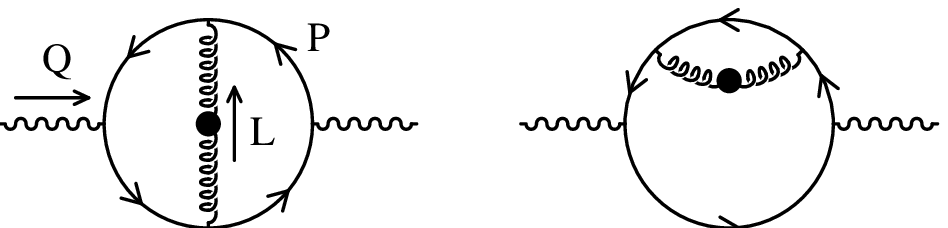}}\hfill \parbox{5.5cm}{\small{{\bf
          Figure\ 7.\ \ } 2-loop diagrams for the photon polarization
        tensor in the HTL effective perturbative expansion.}} \hfill }
\label{fig:2-loop}
\end{figure}
Indeed, these contributions contain bremsstrahlung, and have a large
phase space since the quark loop is hard. The latter property is
enough to compensate the two additional $q\bar{q}g$ vertices. In fact,
it turns out that due to strong collinear singularities regularized by
an asymptotic thermal of order $gT$, the 2-loop contribution is
dominant, and of order ${\rm Im}\,\Pi{}^\mu{}_\mu(Q)\sim e^2 g^2
T^3/q_0$ for soft photons \cite{AurenGKP} and of order ${\rm
  Im}\,\Pi{}^\mu{}_\mu(Q)\sim e^2 g^2 q_0 T$ for hard photons
\cite{AurenGKZ1}. In both cases, the numerical prefactor is of order
$1$ if $Q^2/q_0^2$ is small, but decreases very fast when $Q^2$
increases. In fact, one can check that the collinear enhancement is
controlled by the following quantity \cite{AurenGKP}
\begin{equation}
M_{\rm eff}^2\equiv M^2_\infty +{{Q^2}\over{q_0^2}} p(p+q_0)\; ,
\end{equation}
where $M_\infty\sim gT$ is the asymptotic thermal mass of a hard
quark, and $p$ is the momentum of the quark. The prefactor turns out
to be of order $g^2T^2/M^2_{\rm eff}$.

\subsection{Infrared divergences and KLN cancellations}
Looking carefully at the 2-loop diagrams in Fig.~7, one can check that
the kinematics prevents the gluon momentum $L$ from becoming
arbitrarily small. However, if one considers higher loop diagrams like
the one depicted on Fig.~8, this the argument applies only to the sum
$L_1+L_2$ of the two gluon momenta.  \catcode`\@=11
\global\advance\c@figure by 1 \catcode`\@=12
\begin{figure}[htbp]
  \centerline{\hfill\parbox{2.9cm}{\epsfxsize=2.9cm
      \epsfbox{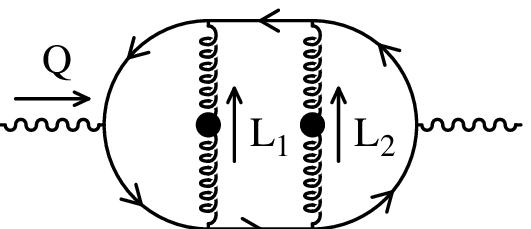}}\hfill \parbox{7.5cm}{\small{{\bf
          Figure\ 8.\ \ }Example of 3-loop contribution to the photon
        polarization tensor.}} \hfill }
\label{fig:3-loop}
\end{figure}
Therefore, one gluon momentum can still go to zero and cause infrared
divergences. One can estimate by power counting the order of magnitude
of this 3-loop contribution \cite{AurenGZ1}, and one readily finds
$3-loop\sim 2-loop\times g^2T/\mu$ where $\mu$ is the infrared cutoff
for the additional (space-like) gluon. If this extra gluon is
longitudinal, then the cutoff is provided by the Debye mass of order
$gT$ so that the 3-loop contribution is suppressed by a power of $g$.
But, if the extra gluon is transverse, then its cutoff can only come
from the non-perturbative magnetic screening mass of order $g^2T$, and
therefore we have $3-loop \sim 2-loop$. This problem is very similar to
the problem found by Linde for the free energy \cite{Linde1}, but
occurs in the leading order for photon production.

However, this power counting argument applies to individual cuts
through the 3-loop diagram, but does not exclude IR cancellations when
one sums the different cuts. In fact, it can be proven \cite{AurenGZ1}
that this sum over the cuts indeed leads to cancellations of infrared
singularities, leaving a finite result without the need of a magnetic
mass, a property which can be seen as a particular case of the
Kinoshita-Lee-Nauenberg theorem \cite{KLN}. More precisely, the sum
over the cuts generates a kinematical cutoff of order $l_{\rm min}\sim
q_0 M^2_{\rm eff}/2p(p+q_0)$, where $M_{\rm eff}$ is the mass
introduced in the 2-loop calculation.

We are now left with two cutoffs that we must compare: the cutoff
$l_{\rm min}$ that comes from the KLN theorem, and an hypothetical
magnetic mass $\mu$ at the scale $g^2T$.  Indeed, it may happen that
$l_{\rm min}$ is smaller than the magnetic mass, and therefore does
not prevent the result from being sensitive to the non-perturbative
scale $g^2T$.  \catcode`\@=11 \global\advance\c@figure by 1
\catcode`\@=12
\begin{figure}[htbp]
  \centerline{\hfill\parbox{6cm}{\epsfysize=3.8cm
      \epsfbox{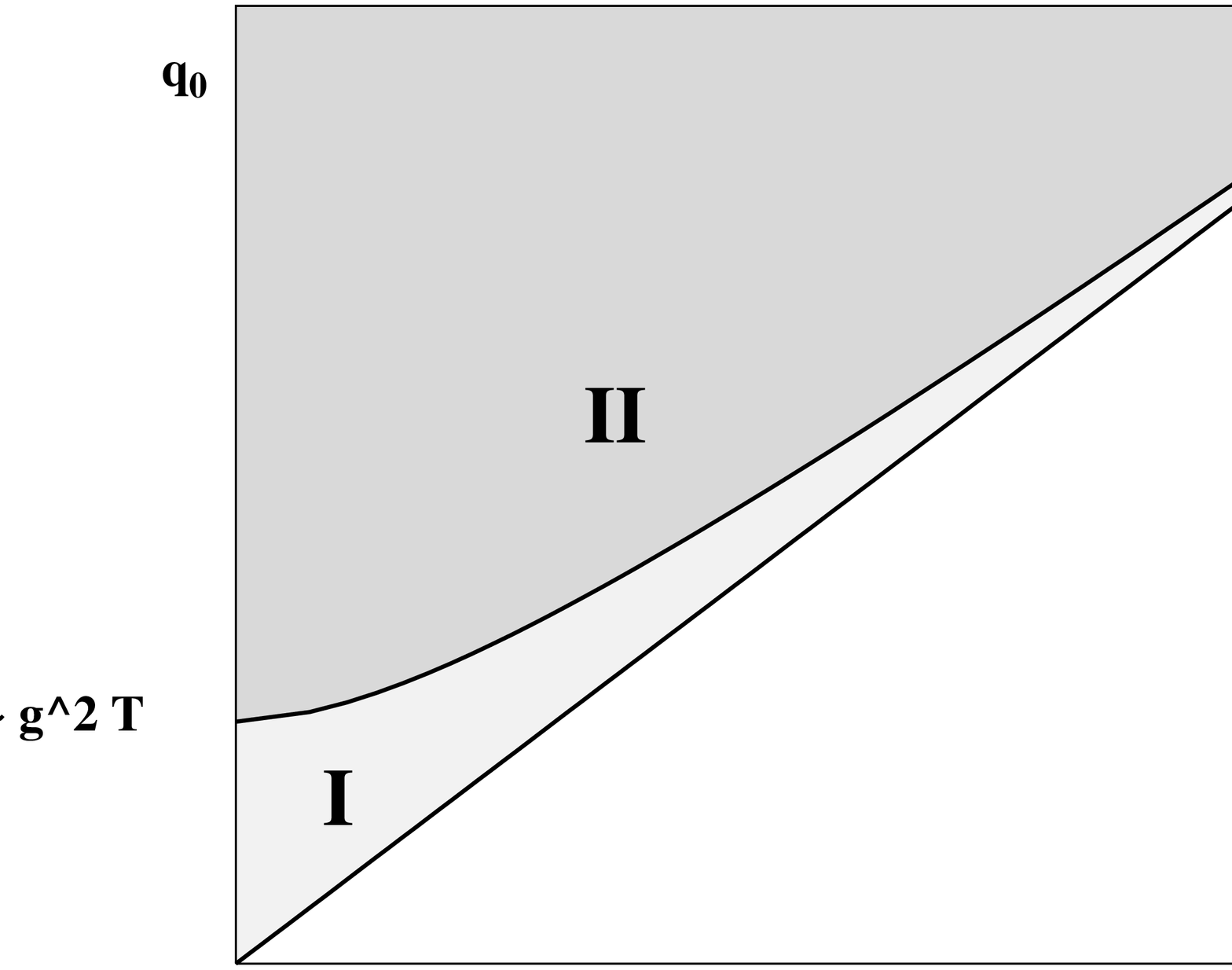}}\hfill
    \parbox{5.5cm}{\small{{\bf Figure\ 9.\ \ }Comparison of $l_{\rm min}$ and
        $\mu\sim g^2T$ in the $(q,q_0)$ plane. In region I, we have
        $\mu> l_{\rm min}$ and the rate is sensitive to the magnetic scale
        despite the KLN cancellations: the photon rate is
        non-perturbative. In region II, we have $\mu< l_{\rm min}$ so that we
        are not sensitive to the magnetic mass. In addition, in region
        II, we have 3-loop$<$2-loop, and the photon rate can be
        treated perturbatively.}} \hfill }
\label{fig:IRsensitivity}
\end{figure}
This comparison has been done in \cite{AurenGZ1} and leads to a
division of the photon phase space in two regions (see Fig.~9): for a
large enough invariant mass (region II), the rate is insensitive to
the scale $g^2T$ and is dominated by the 2-loop contribution, while on
the contrary low invariant mass photons (region I) are sensitive to
the scale $g^2T$ and one must resum an infinite series of ultra-soft
corrections to estimate their rate.

\subsection{Quark lifetime and LPM effect}
At the HTL level of approximation, the quarks have a thermal mass but
their lifetime remains infinite. Indeed, their decay width $\Gamma
\sim g^2T$ appears only at the next order. A particular class of higher
loop corrections which is worth studying by itself is the set of
self-energy corrections that provides a width to the quarks. In fact,
instead of adding perturbatively self-energy corrections on top of the
2-loop diagrams of Fig.~7, it is simpler to use a quark propagator
already including a width. This amounts to perform the following
transformation in the retarded propagator of the quark:
${{(P^2-M^2_\infty)}^{-1}}\to
{{((p_0+i\Gamma)^2-p^2-M^2_\infty)}^{-1}}$.  The calculation with this
propagator is rather straightforward, and leads to results that differ
from the $\Gamma=0$ situation mainly by the expression of $M_{\rm
  eff}^2$:
\begin{equation}
M^2_{\rm eff}=M^2_\infty +{{Q^2}\over{q_0^2}} p(p+q_0)
+4 i{\Gamma\over {q_0}}p(p+q_0)\; .
\end{equation}
Since the width comes in the result via the ratio $\Gamma
p(p+q_0)/q_0$, we see that it affects in an important way the rate of
soft photons, while it modifies only marginally the rate of hard
photons (if $q_0\to \infty$, the imaginary part of $M^2_{\rm eff}$ is
of the same order of magnitude as its real part). Numerical results
for ${\rm Im}\,\Pi{}^\mu{}_\mu(Q)$ are displayed in Fig.~10, where it
is easy to check the above remarks.  \catcode`\@=11
\global\advance\c@figure by 1 \catcode`\@=12
\begin{figure}[htbp]
  \centerline{\hfill\parbox{6cm}{\epsfysize=3.8cm
      \epsfbox{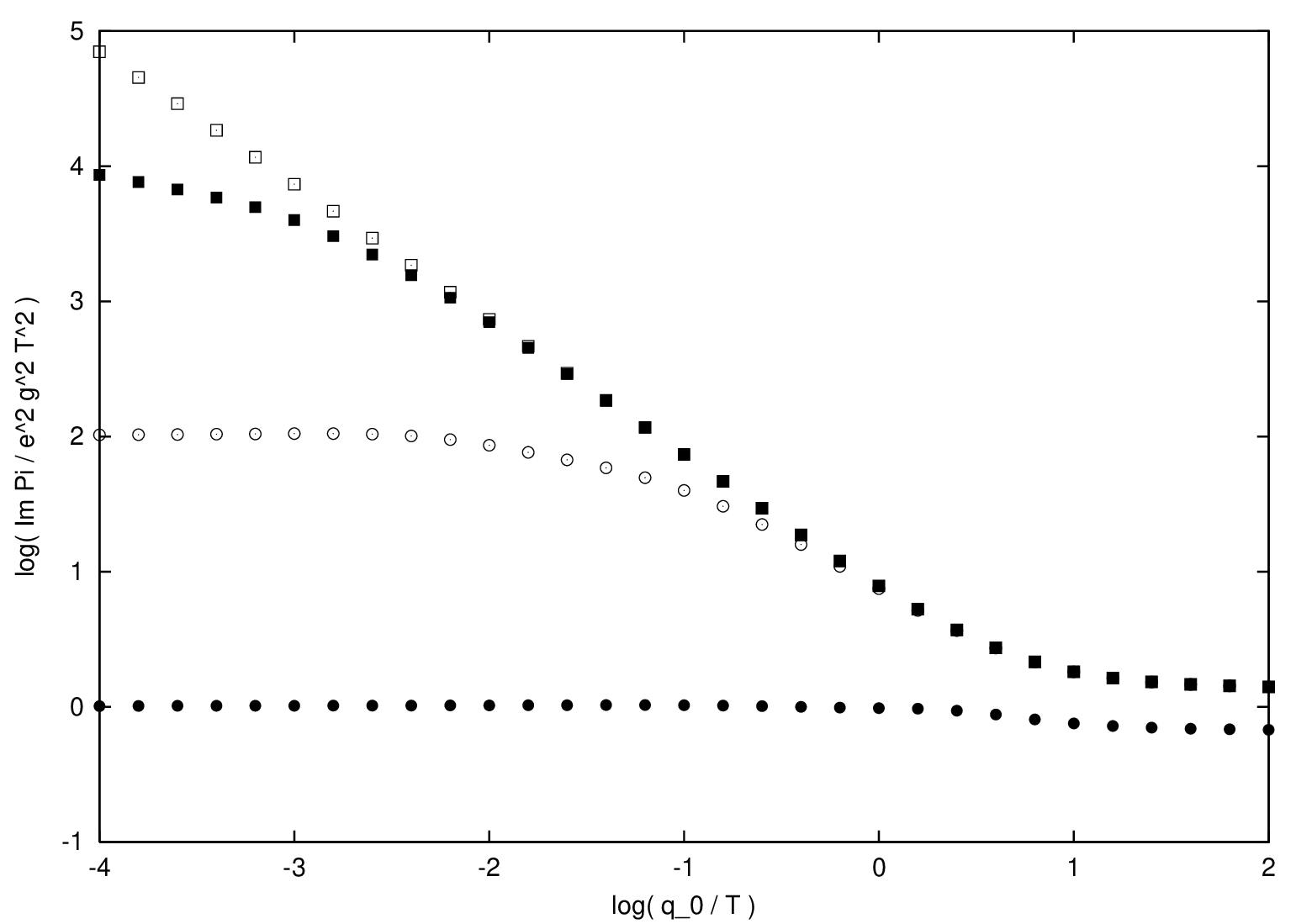}}\hfill \parbox{5.5cm}{\small{{\bf
          Figure\ 10.\ \ }Effect of the width on the bremsstrahlung as
        a function of $q_0/T$ (for $Q^2=0$). The various curves
        correspond to different values of the width $\Gamma$. From top
        to bottom, the ratio $\Gamma T/M^2_\infty$ takes the values
        $10^{-6}$, $10^{-4}$, $10^{-2}$ and $1$.}} \hfill }
\label{fig:width}
\end{figure}

The next step is to determine the region of the photon phase space in
which a width of order $\Gamma\sim g^2T$ would be important. For that
purpose, it is sufficient to compare the real and imaginary part of
the complex number $M^2_{\rm eff}$. They are of the same order of
magnitude when $2\Gamma \sim l_{\rm min}$, where $l_{\rm min}$ is the
kinematical cutoff that appeared in the previous section. This makes
obvious the fact that the region where the width is important is the
same as the region where higher loop corrections are found to be
sensitive to the scale $g^2T$ (region I of Fig.~9). This agreement is
consistent with the fact that the width is dominated by the exchange
of transverse gauge bosons of order $g^2T$ \cite{BlaizI}.

In addition, we can also give a much more physical interpretation for
the non-perturbative region I. Indeed, since $l_{\rm min}$ is the
minimum momentum for an exchanged gluon, its inverse is the coherence
length $\lambda_{\rm coh}$ for the emission (it can also be seen as
the formation time of the photon). The inverse of $\Gamma$ is the mean
free path $\lambda_{\rm mean}$ of the quark in the plasma.  The
inequality $l_{\rm min} < 2\Gamma$ defining region I can therefore be
rewritten as $\lambda_{\rm mean} < \lambda_{\rm coh}$. This condition
is nothing but the criterion for the Landau-Pomeranchuk-Migdal effect
\cite{LPM}. In fact, it is possible to rewrite the parameter $M^2_{\rm
  eff}$ that controls the bremsstrahlung process in a much more
suggestive way:
\begin{equation}
q_0 M^2_{\rm eff}={{2p(p+q_0)}}\left[{{\lambda_{\rm coh}}^{-1}}
+{i{\lambda_{\rm mean}}^{-1}}\right]\; .
\end{equation}
The LPM effect occurs when the formation time of the photon starts to
be larger than the lifetime of the quark: successive scatterings of
the quark cannot be considered individually, and multiple scatterings
modify the production of the photon. In the language of thermal field
theory, the photon production rate becomes sensitive to higher loop
diagrams.

\section{Conclusions}
Concerning the calculation of the QCD pressure, the first remark one
can make is that the best results so far have been obtained in simple
models that just give a mass to the gluons. Despite its lack of
rigorous theoretical basis, this result indicates that most of the
relevant physics can already be grabbed by having realistic
quasiparticles. To include them more rigorously in the formalism,
variational principles derived from the Luttinger-Ward formula seem to
be a good option, but may be very complicated to implement for QCD.
Simplifications can be obtained by restricting the variational space
to the 1-dimensional sub-manifold spanned by HTL corrections, allowing
one to have simultaneously realistic quasiparticles, thermodynamical
consistency, and gauge invariance in a relatively compact formalism.

Concerning photon production by a QGP, it is now clear that the
problem is non-perturbative if the invariant mass of the photon is too
small. Indeed, the necessity of resumming higher order corrections
appeared consistently both in the study of the infrared properties of
higher loop diagrams, and when taking into account the finite timelife
of the quarks. This breakdown of the perturbative expansion for low
mass photons can be interpreted as a manifestation of the LPM effect.
However, finding a practical way to resum {\sl all} the relevant
contributions in this region is still an open question.

\vskip 5mm

\noindent {\bf Acknowledgements.}
It is a pleasure to thank the organizers of the WHEPP'6 workshop
for this stimulating meeting, as
well as the IMSc for its hospitality. My work is supported by DOE
under grant DE-PC02-98CH10886.


\begin{thebibliography}{99}

\bibitem{Frules}
{R.L. Kobes, G.W. Semenoff, N. Weiss}, Z. Phys. {\bf C} {\bf 29}, 371 ({1985}).
{M.A. van Eijck, R. Kobes, Ch.G. van Weert}, Phys. Rev. {\bf D} {\bf 50}, 4097
  ({1994}).

\bibitem{HTL}
{R.D. Pisarski}, Physica {\bf A} {\bf 158}, 146 ({1989}).
{E. Braaten, R.D. Pisarski}, Nucl. Phys. {\bf B} {\bf 337}, 569 ({1990}).
{\sl ibid} {\bf B} {\bf 339}, 310 ({1990}).
{J. Frenkel, J.C. Taylor}, Nucl. Phys. {\bf B} {\bf 334}, 199 ({1990}).
{\sl ibid} {\bf B} {\bf 374}, 156 ({1992}).

\bibitem{BraatN}
{E. Braaten, A. Nieto}, Phys. Rev. Lett. {\bf 76}, 1417 ({1996}).
{\sl ibid} {\bf D} {\bf 53}, 3421 ({1996}).

\bibitem{ArnolZ1}
{P. Arnold, Ch. Zhai}, Phys. Rev. {\bf D} {\bf 51}, 1906 ({1995}).

\bibitem{BoydA1}
{C. Boyd, et Al.}, Nucl. Phys. {\bf B} {\bf 469}, 419 ({1996}).

\bibitem{ParwaS2}
{R. Parwani, H. Singh}, Phys. Rev. {\bf D} {\bf 51}, 4518 ({1995}).

\bibitem{KarscPP1}
{F. Karsch, A. Patkos, P. Petreczky}, Phys. Lett. {\bf B} {\bf 401}, 69
  ({1997}).

\bibitem{PeshiKP1}
{A. Peshier, B. Kampfer, O.P. Pavlenko}, Phys. Rev. {\bf D} {\bf 54}, 2399
  ({1996}).

\bibitem{LuttiW1}
{J.M. Luttinger, J.C. Ward}, Phys. Rev. {\bf 118}, 1417 ({1960}).

\bibitem{PeshiKP2}
{A. Peshier, B. Kampfer, O.P. Pavlenko}, Eur. Phys. Lett. {\bf 43}, 381 ({1998}).

\bibitem{FreedM}
{B.A. Freedman, L. McLerran}, Phys. Rev. {\bf D} {\bf 16}, 1130, 1147, 1169 ({1978}).

\bibitem{QCD-consist}
{J.P. Blaizot, E. Iancu, A. Rebhan}, Phys. Lett. {\bf B} {\bf 470}, 181
  ({1999}).
{A. Peshier}, hep-ph/9910451.

\bibitem{AnderBS}
{J.O. Andersen, E. Braaten, M. Strickland}, Phys. Rev. Lett. {\bf 83}, 2139
  ({1999}).
{\sl ibid}, Phys. Rev. {\bf D} {\bf 61}, 014017
  ({2000}).

\bibitem{rate}
{H.A. Weldon}, Phys. Rev. {\bf D} {\bf 28}, 2007 ({1983}).
{C. Gale, J.I. Kapusta}, Nucl. Phys. {\bf B} {\bf 357}, 65 ({1991}).

\bibitem{soft-gamma}
{R. Baier, S. Peign{\accent 19 e}, D. Schiff}, Z. Phys. {\bf C} {\bf 62}, 337
  ({1994}).
{P. Aurenche, T. Becherrawy, E. Petitgirard}, hep-ph/9403320 (unpublished) .
{A. Niegawa}, Phys. Rev. {\bf D} {\bf 56}, 1073 ({1997}).

\bibitem{hard-gamma}
{R. Baier, H. Nakkagawa, A. Niegawa, K. Redlich}, Z. Phys. {\bf C} {\bf 53},
  433 ({1992}).
{J.I. Kapusta, P. Lichard, D. Seibert}, Phys. Rev. {\bf D} {\bf 44}, 2774
  ({1991}).

\bibitem{FlechR1}
{F. Flechsig, A.K. Rebhan}, Nucl. Phys. {\bf B} {\bf 464}, 279 ({1996}).

\bibitem{AurenGKP}
{P. Aurenche, F. Gelis, R. Kobes, E. Petitgirard}, Phys. Rev. {\bf D} {\bf 54},
  5274 ({1996}).
{\sl ibid}, Z. Phys. {\bf C} {\bf 75},
  315 ({1997}).

\bibitem{AurenGKZ1}
{P. Aurenche, F. Gelis, R. Kobes, H. Zaraket}, Phys. Rev {\bf D} {\bf 58},
  085003 ({1998}).

\bibitem{AurenGZ1}
{P. Aurenche, F. Gelis, H. Zaraket}, hep-ph/9911367, to appear in Phys. Rev. D.

\bibitem{Linde1}
{A.D. Linde}, Phys. Lett. {\bf B} {\bf 96}, 289 ({1980}).

\bibitem{KLN}
{T.D. Lee, M. Nauenberg}, Phys. Rev. {\bf 133}, 1549 ({1964}).
{T. Kinoshita}, J. Math. Phys. {\bf 3}, 650 ({1962}).

\bibitem{BlaizI}
{J.P. Blaizot, E. Iancu}, Nucl. Phys. {\bf B} {\bf 459}, 559 ({1996}).
{\sl ibid}, Phys. Rev. Lett. {\bf 76}, 3080 ({1996}).

\bibitem{LPM}
{L.D. Landau, I.Ya. Pomeranchuk}, Dokl. Akad. Nauk. SSR {\bf 92}, 535, 735 ({1953}).
{A.B. Migdal}, Phys. Rev. {\bf 103}, 1811 ({1956}).

\end{thebibliography}

\end{document}